%% file: aswec2015.tex
\documentclass[10pt, conference, compsocconf]{IEEEtran}

\usepackage[cmex10]{amsmath}
\usepackage{cite}
\usepackage{times}
\usepackage{xspace}
\usepackage{ifthen}
\usepackage{amssymb}
\usepackage{caption}

\usepackage{numprint}
\npthousandsep{,}        %% could be changed to another symbol

\usepackage{graphicx}
\graphicspath{{figures/}}

\usepackage{framed}

\newboolean{showcomments}
\setboolean{showcomments}{true} % TOGGLE true/false to show or hide comments

\ifthenelse{\boolean{showcomments}}
  {\newcommand{\note}[2]{
	\fbox{\bfseries\sffamily\scriptsize#1}
    {\sf\small$\blacktriangleright$\textit{#2}$\blacktriangleleft$}
   }

  }
  {\newcommand{\note}[2]{}
   
  }

\newcommand{\gap}{{\boldsymbol \star}}
\newlength{\aligncharw}
\newcommand{\agap}{\makebox[\aligncharw][s]{\scriptsize{$\gap$}}}

\newcommand{\eg}{\emph{e.g.}\xspace}
\newcommand{\etal}{\emph{et al.}\xspace}
\newcommand{\ie}{\emph{i.e.}\xspace}
\newcommand{\cf}{\emph{cf.}\xspace}

\newcommand{\bigleftmost}{\emph{biggest-left-most-common-subword}\xspace}

\newcommand{\MS}{\emph{Mandile-Schneider}\xspace}

\makeatletter
\newcommand\footnoteref[1]{\protected@xdef\@thefnmark{\ref{#1}}\@footnotemark}
\makeatother

\begin{document}

\title{Generalized Suffix Tree based Multiple Sequence Alignment for Service Virtualization}

%%% Submitted title %%%
%% \title{Modeling and Validating Enterprise System Protocol Specifications}

%%% Old Title %%%
%% \title{Modeling Enterprise System Protocol Implementations and Verifying Trace Conformance}

\author{\IEEEauthorblockN{Jean-Guy Schneider and Peter Mandile}
\IEEEauthorblockA{Faculty of Science, Engineering and Technology\\
Swinburne University of Technology\\
P.O. Box 218,
Hawthorn, VIC 3122, AUSTRALIA \\
{\tt \{jschneider,pmandile\}@swin.edu.au}}
\and
\IEEEauthorblockN{Steve Versteeg}
\IEEEauthorblockA{CA Labs \\
Level 2, 380 St. Kilda Rd \\
Melbourne, VIC 3004, AUSTRALIA \\
{\tt steven.versteeg@ca.com}}
}

\maketitle

% ------------------------------------------------------------------------- %

\input{abstract}


\IEEEpeerreviewmaketitle

\input{introduction}

%%% add more sections here...
\input{background}
\input{preliminaries}

\input{algorithm}
\input{evaluation}
\input{conclusions}

% ------------------------------------------------------------------------- %

%% \bibliographystyle{ieeetr}

%%% Old IEEE Formatting (CMH)
%\bibliographystyle{ieee}

\bibliographystyle{IEEETran}
\bibliography{./aswec2015}

% ------------------------------------------------------------------------- %

\end{document}

%% file: abstract.tex
% Document Type: LaTeX
% Master File: abstract.tex
% Author: Jean-Guy Schneider, Peter Mandile
% Last Modified: Fri Aug 28 2015, 14:35

% Abstract for the ASWEC 2015 submissions

%%%%%%%%%%%%%%%%%%%%%%%%%%%%%%%%%%%%%%%%%%%%%%%%%%%%%%%%%%%%%%%%%%%%%%%%%%%%

\begin{abstract}
  Assuring quality of contemporary software systems is a very challenging task
  due to the often large complexity of the deployment environments in which
  they will operate. Service virtualization is an approach to this challenge
  where services within the deployment environment are emulated by
  synthesising service response messages from models or by recording and then
  replaying service interaction messages with the system. Record-and-replay
  techniques require an approach where (i) message prototypes can be derived
  from recorded system interactions ({\ie} request-response sequences), (ii) a
  scheme to match incoming request messages against message prototypes, and
  (iii) the synthesis of response messages based on similarities between
  incoming messages and the recorded system interactions. Previous approaches
  in service virtualization have required a multiple sequence alignment (MSA)
  algorithm as a means of finding common patterns of similarities and
  differences between messages required by all three steps.

  In this paper, we present a novel MSA algorithm based on Generalized Suffix
  Trees (GSTs). We evaluated the accuracy and efficiency of the proposed
  algorithm against six enterprise service message trace datasets, with the
  proposed algorithm performing up to 50 times faster than standard MSA
  approaches. Furthermore, the algorithm has applicability to other domains
  beyond service virtualization.

%  Assuring quality of contemporary software systems is a very challenging task
%  due to the often large complexity of the deployment environments in which
%  they will operate. Service virtualization is an approach to this challenge
%  where services within the deployment environment are emulated by
%  synthesising service response messages from models or by recording and then
%  replaying service interaction messages with the system. Record-and-replay
%  techniques require an approach where (i) message prototypes can be derived
%  from recorded system interactions (i.e. request-response sequences), (ii) a
%  scheme to match incoming request messages against message prototypes, and
%  (iii) the synthesis of response messages based on similarities between
%  incoming messges and the recorded system interactions.
%
%  In this paper, we present a novel Multiple Sequence Alignment (MSA)
%  algorithm based on Generalized Suffix Trees (GSTs) that allows for the
%  identification of similarities and differences between messages required by
%  all three steps of record-and-replay techniques. The accuracy and
%  effectiveness of the approach is demonstrated using predefined interaction
%  sequences of a number of enterprise-level protocols. The results stipulate
%  that the algorithm has a broader applicability to other domains beyond 
%  service virtualisation.
%

\end{abstract}

%%%%%%%%%%%%%%%%%%%%%%%%%%%%%%%%%%%%%%%%%%%%%%%%%%%%%%%%%%%%%%%%%%%%%%%%%%%%

%%% Previous ICSE abstract...

% Testing enterprise software systems is very challenging due to the large
% complexity of their eventual deployment environments in which they will
% operate. Common approaches are to emulate services within the deployment
% environment by synthesising service response messages from models or by
% recording and then replaying service interaction messages with the
% system. Models require deep knowledge of the target services while the
% record-and-replay approaches are limited in accuracy and efficiency. In this
% paper, we present a new technique that significantly improves the accuracy of
% the record-and-replay approaches, and does not require any prior knowledge of
% the services (models). It uses a multiple sequence alignment algorithm to
% derive message prototypes from recorded system interaction (request-response)
% message sequences or traces, and a scheme to match incoming request messages
% against message prototypes and generate corresponding response messages. In
% particular, we introduce wildcards in message prototypes for the sections with
% high variability, a modified Needleman-Wunsch algorithm for distance
% calculation during message matching, and entropy-based weightings in distance
% calculations to increase accuracy. Combined, our new approach has shown
% greater than 99\% accuracy for the 4 enterprise system messaging protocols for
% which we have evaluated it.

%%%%%%%%%%%%%%%%%%%%%%%%%%%%%%%%%%%%%%%%%%%%%%%%%%%%%%%%%%%%%%%%%%%%%%%%%%%%

%% file: introduction.tex
% Document Type: LaTeX
% Master File: introduction.tex
% Author: Jean-Guy Schneider, Peter Mandile, Steve Versteeg
% Last Modified: Fri Aug 28 2015, 14:42

% Introduction for the ASWEC 2015 submissions

%%%%%%%%%%%%%%%%%%%%%%%%%%%%%%%%%%%%%%%%%%%%%%%%%%%%%%%%%%%%%%%%%%%%%%%%%%%%

\section{Introduction}
\label{intro.sec}

%%% Basic statement of context -> may need some tweaking...
Organizations today are more reliant than ever on an IT infrastructure to
deliver their services, and the software systems providing these services are
increasingly interconnected and interdependent. Ensuring quality and correct
interoperation of these systems is paramount in achieving uninterrupted
business operations in these organizations. However, these software systems
are mainly developed in isolation without ready access to testing environments
that truly reflect the complexity of real-world deployment environments. As a
consequence, we are faced with the increasing risk that unidentified flaws
cause a cascade of failures across multiple systems, bringing down an entire
IT infrastructure and causing severe interruptions to business operations.

%%% Service virtualization and what it needs
One popular approach that developers use to test their application's
dependence on other systems is to install the other systems on virtual
machines \cite{sugerman:01a}. However, virtual machines are, in general, time
consuming to configure and maintain, and the configuration of the systems
running on virtual machines is likely to be different to the real deployment
environments. An alternative approach is {\em service emulation} or {\em
  service virtualization} where models of services are emulated -- sometimes
into the many thousands of service instances -- to provide more realistic
scale and less complicated configuration \cite{hine:09a}. However, service
emulation often relies on system experts explicitly modelling the target
services and hence require detailed knowledge of the underlying message
protocols and structure, respectively. This is often infeasible if the
required knowledge is unavailable.

%%% Multiple Sequence Alignment -> brief context
We have been working on an automated approach to service emulation which uses
no explicit knowledge of the services, their message protocols and structures,
but solely relies on recordings of interactions between a system under test
and its environment services \cite{du:13a,du:13b}. One key aspect of the
approach is to identify similarities between messages and exploit these
similarities in response generation. Commonly used {\em multiple sequence
  alignment} (MSA) techniques from biology \cite{bieganski:94a,thompson:94a}
were used to do so -- refer to \cite{du:13b} for the details of the approach.

%%% Motivation for novel algorithm
However, our experiments have shown that for larger clusters of messages, the
memory required to perform a multiple sequence alignment as well as the
corresponding computation time increased significantly. Whereas in
bioinformatics the main concern is to align few (generally less than 10), but
{\em long} sequences of amino acids, our approach requires the alignment of
many (100s to thousands), but {\em shorter} messages. MSA algorithms
optimized for the biological use case may therefore be unsuited for the
service emulation domain.

The main contribution of this paper is a novel, memory and time efficient MSA
algorithm based on {\em Generalized Suffix Trees} \cite{weiner:73a} that
addresses these shortcomings. The efficiency and accuracy of the algorithm is
evaluated using a set of experiments with typical enterprise system messaging
protocols, including LDAP, a binary mainframe protocol, and SOAP services, and
compare it with the standard MSA algorithm used in our prior work.

%%% Overview Section
\smallskip

The rest of this paper is organized as follows: we further motivate our work
on a novel multiple sequence alignment (MSA) algorithm suitable for the
service emulation domain in Section~\ref{background.sec}, followed by a
discussion of other MSA algorithms in Section~\ref{msa.sec}. In
Section~\ref{msalgo.sec}, we present our novel, GST-based multiple sequence
alignment approach. The results of our experiments in evaluating the
effectiveness and accuracy of the new approach are presented and discussed in
Section~\ref{eval.sec}. Finally, in Section \ref{conc.sec}, we summarize the
main findings of this paper and give an outlook for future work.

%%%%%%%%%%%%%%%%%%%%%%%%%%%%%%%%%%%%%%%%%%%%%%%%%%%%%%%%%%%%%%%%%%%%%%%%%%%%

%% file: background.tex
% Document Type: LaTeX
% Master File: background.tex
% Author: Jean-Guy Schneider, Peter Mandile, Steve Versteeg
% Last Modified: Fri Aug 28 2015, 14:49

% Background Section for the ASWEC 2015 submissions

%%%%%%%%%%%%%%%%%%%%%%%%%%%%%%%%%%%%%%%%%%%%%%%%%%%%%%%%%%%%%%%%%%%%%%%%%%%%

\section{Background}
\label{background.sec}

The main goal of our work is to produce an emulation environment for
enterprise system testing that does not rely on a priori knowledge of the
message protocols and structure used by the environment services, but uses
message trace recordings collected a priori to produce a response on behalf of
a service when invoked by a system under test (SUT) at run-time.

The main idea behind our approach is that if a request sent to an environment
service by a SUT is very similar to a recorded request for this service
(having a suitable notion of ``similarity''), then the response to this
request is expected to be similar to the corresponding previously recorded
response. Hence, identifying the differences between the incoming and
previously recorded requests should give us a good indication how the
corresponding recorded response can be altered in order to synthesize a
matching response.

In \cite{du:13a}, we used an approach where an incoming message was compared
with {\em all} recorded messages in order to determine the most similar one.
We used a normalized {\em edit distance} \cite{ristad:98a} as the similarity
measure. Although this approach produced very accurate responses, it does not
scale well to large transaction libraries. Using a dynamic programming
approach, the edit distance between two messages $m_1$ and $m_2$ with
corresponding lengths $m_1^l$ and $m_2^l$ has a time complexity of
$\mathcal{O}(m_1^l m_2^l)$. Hence, if we have a transaction library with $n$
recorded messages and an average message length of $\overline{m}$, then
finding the most similar recorded request has a time complexity of
$\mathcal{O}(n^2 \overline{m}^2)$.

In order to improve efficiency, we extended to original approach by clustering
the trace recordings into groups of similar messages (ideally of the same {\em
  operation type}) and then formulate a single representation for the request
messages for each cluster \cite{du:13b}. This accelerated run-time performance
by enabling incoming requests from the system under test to be compared only
to the cluster representations, rather than the entire transaction
library. However, choosing the cluster \emph{centroid}\footnote{The centroid
  is the transaction with the minimized total distance from the other
  transactions in the cluster.} request as the representative resulted in a
decreased accuracy of the generated responses as the information from the
other requests in the cluster was discarded.

Therefore, we need an approach that addresses both concerns, efficiency and
accuracy, by generating cluster \emph{prototypes} which capture the common
features of the range of requests in each cluster, but also preserves their
variability as much as possible.

\smallskip

%%% Regular expressions...
In general, messages of typical enterprise-level protocols contain a mix of
structural information (which is mostly {\em identical} for the same operation
type) as well as payload information that varies from message to message.
Hence, it should be possible to ``summarize'' the requests within each cluster
using a suitably crafted {\em regular expression} \cite{sudkamp:05a}
containing constant strings for structural information and patterns for
payload information, respectively.
As an example, consider the messages in the search cluster given in Table
\ref{sampleops.tab}. They all conform to the regular expression

\smallskip
{\centerline {
{\small {\tt {\{id:[1-9][0-9]*,op:S,sn:[A-Z][a-z]+\} }}}
}}

\noindent
assuming that message identifiers do not start with a '$0$' and can have an
arbitrary length and all search strings start with an uppercase character.
Similarly, all messages of the add cluster conform to a slightly more complex
regular expression. One of the key advantages of regular expressions is that
they generally match a broader range of messages than just the ones they were
extracted from, as long as these message follow the same basic message
structure.

% {\small {\tt {\{id:[1-9][0-9]*,op:A,sn:[A-Z][a-z]+(,[A-Z]*[a-z]+:[A-Za-z0-9]+)*\} }}}

Initial experiments with manually constructing cluster prototypes for LDAP
\cite{sermersheim:06a} showed that indeed, all clusters prototypes could be
expressed as regular expressions. This leads us to the question how to best
extract a regular expression from a given set of input sequences.

% Initial experiments with manually constructing a single regular expression as
% the cluster prototype for textual LDAP [REF and/or hyperlink needed] not only
% showed that it was indeed possible to create regular expressions using the
% strcuture mentioned above, but also demonstrated the effectiveness of
% identifying the correct cluster for an incoming message by comparing it
% against all cluster prototypes.

%%%%%%%% The 'fake' messages for demonstration purposes
\begin{table}[t]
\begin{center}
\begin{tabular}{|c|l|}
\hline
Cluster & Request Message \\
\hline\hline
       & \{id:1,op:S,sn:Smith\} \\
       & \{id:275,op:S,sn:Miller\} \\
Search & \{id:13,op:S,sn:Wilson\} \\
       & \{id:2273,op:S,sn:Mandile\} \\
       & \{id:490,op:S,sn:Schneider\} \\
\hline
    & \{id:24,op:A,sn:Schneider,mobile:123456\} \\
Add & \{id:4287,op:A,sn:William\} \\
    & \{id:3206,op:A,sn:Turner,gn:Samuel,Postcode:34589\} \\
\hline
\end{tabular}
\end{center}
\caption{Sample search and add operations}
\label{sampleops.tab}
\end{table}

\smallskip

Existing approaches for the generation of regular expressions from a input
samples can be classified into two broad categories: either they generate a
finite automaton for each input sequence and then apply a number of rules to
merge the individual automata into a single automaton, or they use a multiple
sequence alignment (MSA) based approach.

%%% Some related work here... maybe will go elsewhere...
Watson's approach \cite{watson:03a}, for example, falls into the first
category. He presented a semi-incremental algorithm for constructing minimal
acyclic deterministic finite automata for a given set of sequences. These
automata can then be translated into regular expressions. However, the
generated automata only exploit common parts at the beginning and end of the
input sequences (in our search cluster example, `{\small {\tt {{\{id:}}}}' and
`{\small {\tt {\}}}}'), but not in-between, resulting in rather complex
automata. Also, the resulting automata only accept sequences from the given
input, but do not accept any sequences not covered by the input, which is of
rather limited use for our purpose.

%%% I need to find the reference for this. Otherwise, not point mentioning
%%% it!!!
% Blah [REF - need to check it!] presented an approach for regular expression
% generation based on a progressive pair-wise sequence alignment
% However, the algorithm used to align the sequences is not optimal, and any
% non-constant part is summarized using the pattern {\sf '.*'}.

%%% Using MSA to generate the necessary input for REs
MSA-based approaches, such as for example the one presented by Tang {\etal}
\cite{tang:07a}, perform an alignment of all sequences, identify the
overlapping (or aligned) sub-sequences as constant strings of the resulting
regular expression, and extract suitable patterns for the gaps in-between.
Consider the multi-sequence alignment of the messages in the search cluster
given in Table \ref{sampleops.tab} (with {$\gap$} denoting the {\em gap
  symbol} introduced during the alignment):

%%% Aligned search sequences... the corresponding macro...
\newcommand{\alignedSrqs}{
\begin{minipage}[b]{.45\linewidth}
\texttt{\footnotesize{
\{id:1{\agap}{\agap}{\agap},op:S,sn:Smith{\agap}{\agap}{\agap}{\agap}\} \linebreak
\{id:275{\agap},op:S,sn:Miller{\agap}{\agap}{\agap}\} \newline
\{id:13{\agap}{\agap},op:S,sn:Wilson{\agap}{\agap}{\agap}\} \newline
\{id:2273,op:S,sn:Mandile{\agap}{\agap}\} \newline
\{id:490{\agap},op:S,sn:Schneider\} \newline
}}
\end{minipage}
}

\noindent {\centerline {\alignedSrqs}}

\noindent The alignment results in three overlapping sub-sequences ({\ie}
`{\small {\tt {{\{id:}}}}', `{\small {\tt {,op:S,sn:}}}' and `{\small {\tt
    {\}}}}'), the same constant sub-sequences that were used to form the
regular expression shown above.

\smallskip

For the rest of this work, we solely focus on identifying the overlapping (or
aligned) sub-sequences of a set of messages in order to identify the constant
parts of regular expression-based cluster prototypes. Finding suitable
patterns for the gaps between the constant strings that appropriately reflect
the nature of the payload is a topic of an ongoing investigation.

%%%%%%%%%%%%%%%%%%%%%%%%%%%%%%%%%%%%%%%%%%%%%%%%%%%%%%%%%%%%%%%%%%%%%%%%%%%%

%% file: preliminaries.tex
% Document Type: LaTeX
% Master File: preliminaries.tex
% Author: Jean-Guy Schneider, Peter Mandile, Steve Versteeg
% Last Modified: Fri Aug 28 2015, 15:05

% Mathematical Preleminaries for the ASWEC 2015 submissions

%%%%%%%%%%%%%%%%%%%%%%%%%%%%%%%%%%%%%%%%%%%%%%%%%%%%%%%%%%%%%%%%%%%%%%%%%%%%

%%% MACROS for the formalisms... not all used any more...
\newcommand{\word}{word}
\newcommand{\words}{words}
\newcommand{\subword}{sub-word}
\newcommand{\suffix}{suffix}
\newcommand{\length}{length}
\newcommand{\csw}{common sub-word}
\newcommand{\msw}{multi sub-word}

\newcommand{\CharSet}{\mbox{$\mathcal{C}$}}
\newcommand{\WordSet}{\mbox{$\mathcal{W}$}}
\newcommand{\SubWord}[3]{\mbox{$(#1,#2,#3)$}}

\newcommand{\MSW}[2]{[ {#1} -- \{ #2 \} ]}
\newcommand{\CSW}[2]{( #1 -- [ #2 ] )}

% ------------------------------------------------------------------------ %

\section{Suffix Trees}
\label{prelim.sec}

In this section, we will briefly introduce a number of concepts required to
illustrate the proposed new multi-sequence alignment algorithm, especially the
notion of a {\em Generalized Suffix Tree} (GST) our approach is heavily based
upon.

% ------------------------------------------------------------------------ %

\subsection{Sequences and Suffixes}

We start with the notion of the most basic building block for our study, the
set of {\em characters}, denoted by {\CharSet}. We require equality and
inequality to be defined for the elements of {\CharSet}. For the purpose of
our study, {\CharSet} will most likely comprise of the set of valid Bytes that
can be transmitted over a network or the set of printable Characters as a
dedicated subset. A word (or {\em sequence}) $s$ is a non-empty, finite
sequence of characters $c_0c_1c_2\ldots c_{n-1}$ with $c_i \in \CharSet, \: 0
\leq\! i\! <\! n$. The {\em {\length}} of a sequence $s$, denoted by $s^l$,
is equal to the number of characters, that is, $s^l = n$. Throughout the rest of
this section we will use the terms word and sequence interchangeably.

%%% Sub-word and suffix
Given a word $s \: = \: c_0c_1c_2\ldots c_{n-1}$, we define a {\em {\subword}}
as a non-empty sub-sequence of characters from $s$ starting at index $i$ and
having length $l$ as $sw_s (i,l) = c_ic_{i+1}c_{i+2}\ldots c_{i+l}$. We
require $l > 0$ and $i\!+\!l \leq n$. A special kind of subword is a {\em
  suffix}, a sub-sequence $c_ic_{i+1}c_{i+2}\ldots c_{n-1}$ that ends with the
last character $c_{n-1}$ of the sequence $s$.

% ------------------------------------------------------------------------ %

\subsection{Suffix Tree}

%%% Image of the suffix tree for 'Banana'
\begin{figure}[t]
\centering
\includegraphics[width=0.80\columnwidth]{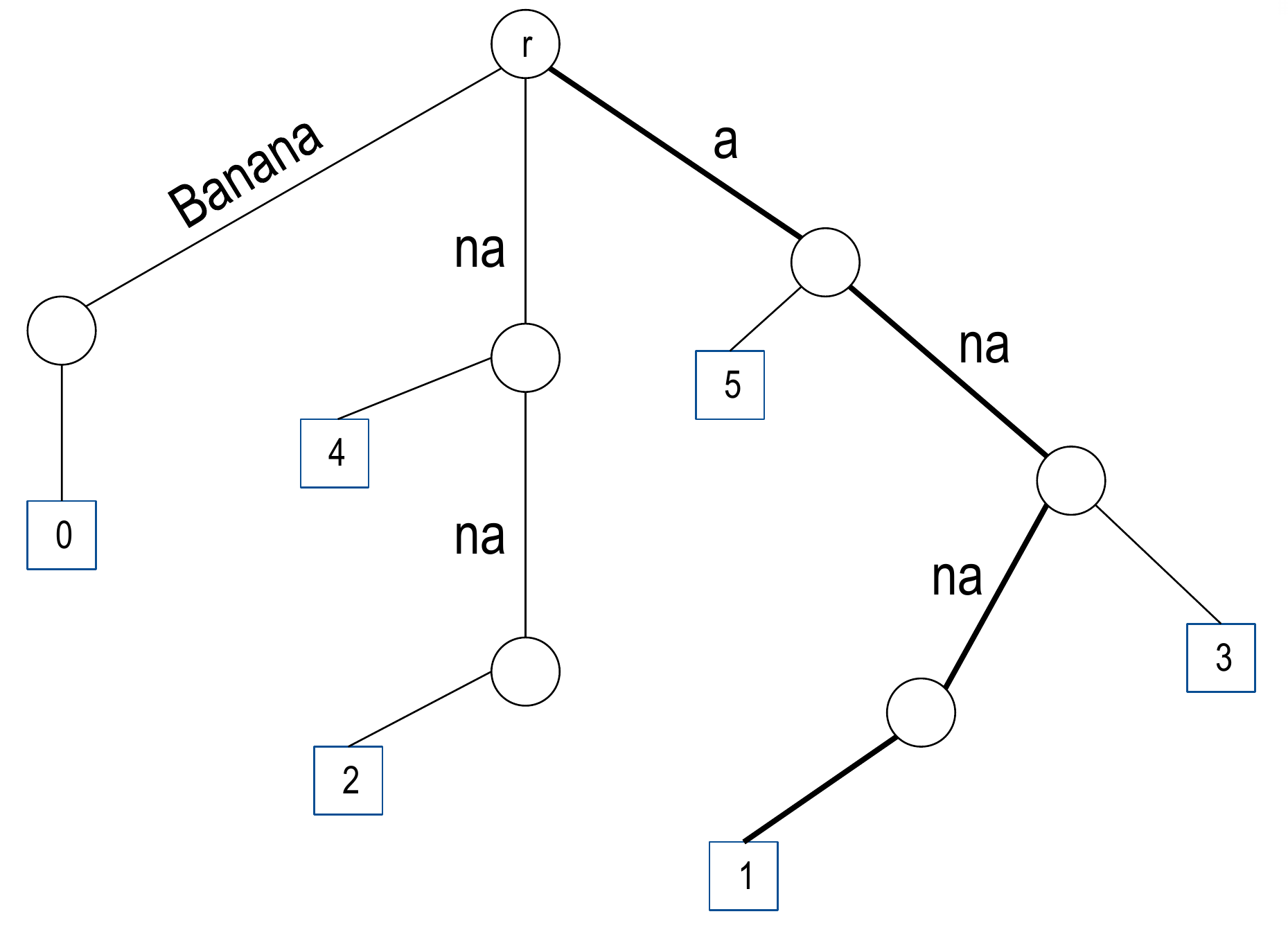}
\caption{Suffix Tree for the word {\em Banana}.}
\label{banana.fig}
\end{figure}

A suffix tree for a sequence $s$, denoted by $ST(s)$, is a tree that contains
all {\em suffixes} of $s$, that is, the sub-sequences $c_{n-1}$,
$c_{n-2}c_{n-1}$ etc. \cite{weiner:73a}. More specifically, a suffix tree is an
{\em edge-labelled tree} containing (i) $n$ labeled leaf nodes and (ii) up to
$n\!+\!1$ non-leaf (or {\em branching}) nodes. The concatenation of the edge
labels of the path from the root node of the tree to the leaf node with label
$i$ defines the suffix $c_ic_{i+1}\ldots c_{n-1}$.

In order to minimize the number of branching nodes in the tree, a branching
node $b$ contains {\em at most} one outgoing edge labelled $c\alpha$ for a
given character $c \in \CharSet$ and some (possibly empty) sequence $\alpha$
with both, $\alpha$ and $c\alpha$ being {\subword}s of the sequence $s$. The
edges to leaf nodes do not contain any labels.{\footnote{Some references
    ({\eg} \cite{ukkonen:94a,hoehl:02a}) use a special character $\$$ to label
    edges to leaf nodes, indicating a dedicated {\em end-of-word}
    character. For the purpose of this presentation, we have omitted edge
    labels to leaf nodes.}} Similarly, the incoming edge to a branching node
$b \neq r$ must contain a non-empty label $\beta$ such that $\beta$ is a
{\subword} of $s$ and for each outgoing edge of $b$ to a branching node $b'$
(with edge label $\gamma$), $\beta\gamma$ is also a {\subword} of $s$.
In order to guarantee that a suffix tree for a sequence with $n$ characters
has no more than $n\!+\!1$ branching nodes, each branching node must have
either an outgoing edge to at least one leaf node or to two branching nodes,
respectively.

Consider the suffix tree for {\em Banana} as given in Figure~\ref{banana.fig}.
Branching nodes are depicted as circles, leaf nodes as squares. The suffix
tree has a uniquely identifiable root node, denoted by $r$, $6$ further
branching nodes, and $6$ leaf nodes (labelled $0$ to $5$). The path
highlighted in Figure~\ref{banana.fig} (from $r$ to leaf node with label $1$)
denotes the suffix starting at index $1$, that is, `{\em anana}'.

It is well known that suffix trees can be constructed in linear time and with
linear space \cite{weiner:73a,ukkonen:94a}.

% ------------------------------------------------------------------------ %

\subsection{Generalized Suffix Tree}

A {\em generalized suffix tree} (or short GST) generalizes the concept of a
suffix tree to more than one sequence, that is, it contains all suffixes for
sequences $s_0, s_1, \ldots, s_{m-1}$ (with $m > 1$). In order to distinguish
which sequence a particular suffix belongs to, leaf nodes contain two labels:
one for the sequence, and one for the start index of the suffix in the
corresponding sequence. As a consequence, a GST for the sequences $s_0, s_1,
\ldots, s_{m-1}$ contain $s^l_0 + s^l_1 + \ldots + s^l_{m-1}$ leaf nodes. The
other properties as given for a suffix tree still hold. A GST can be
constructed in $\mathcal{O} (s_0^l + s_1^l + \ldots s_{m-1}^l)$ time 
and space, respectively \cite{ukkonen:94a}.

%%% Image of the suffix tree for 'DCxzDCxBAQ' and 'DCxDCDCxpxBAQ'
% \begin{figure}[t]
% \centering
% \includegraphics[width=\columnwidth]{xBAQ2}
% \caption{Generalized Suffix Tree for \texttt{DCxzDCxBAQ} and
%   \texttt{DCxDCDCxpxBAQ}.}
% \label{xBAQ2.fig}
% \end{figure}

%%% colour set
We assign a {\em colour set} to each branching node in a GST \cite{chi:92a}.
Each sequence has a unique colour, and we assign this colour to all leaf nodes
that represent a suffix for this sequence. The colour set of each branching
node is the union of the colour set(s) of all of its children. As a consequence,
the colour set of the root node $r$ contains the colours of all sequences. The
colour set of a branching node can be determined in a top-down tree
construction process and does not need to be re-computed bottom-up once a GST
is fully constructed. Hence, the computation of the colour sets does not add
to the algorithmic complexity of the tree construction.

The reader may note that from a sequence alignment perspective, of special
interest are branching nodes that have the same colour set as the root node
\cite{chi:92a}. We denote such branching nodes as {\em fully coloured}. The
concatenation of the edge labels of the paths from $r$ to these nodes
correspond to sub-sequences that are {common} to {\em all} sequences of the
GST.

%%% Old example %%%
% As an example, consider the generalized suffix tree for \texttt{DCxzDCxBAQ}
% and \texttt{DCxDCDCxpxBAQ} as given in Figure~\ref{xBAQ2.fig}. In order to
% enhance readability, the leaf nodes for \texttt{DCxzDCxBAQ} and
% \texttt{DCxDCDCxpxBAQ} are denoted by blue and green squares,
% respectively. All fully coloured nodes are highlighted in red. As we can infer
% from Figure~\ref{xBAQ2.fig}, the two sequences contain the common
% sub-sequences \texttt{Q}, \texttt{AQ}, \texttt{BAQ}, \texttt{x},
% \texttt{xBAQ}, \texttt{C}, \texttt{Cx}, \texttt{CxBAQ}, \texttt{DC}, and
% \texttt{DCx}.

\begin{figure}[t]
\centering
\includegraphics[width=\columnwidth]{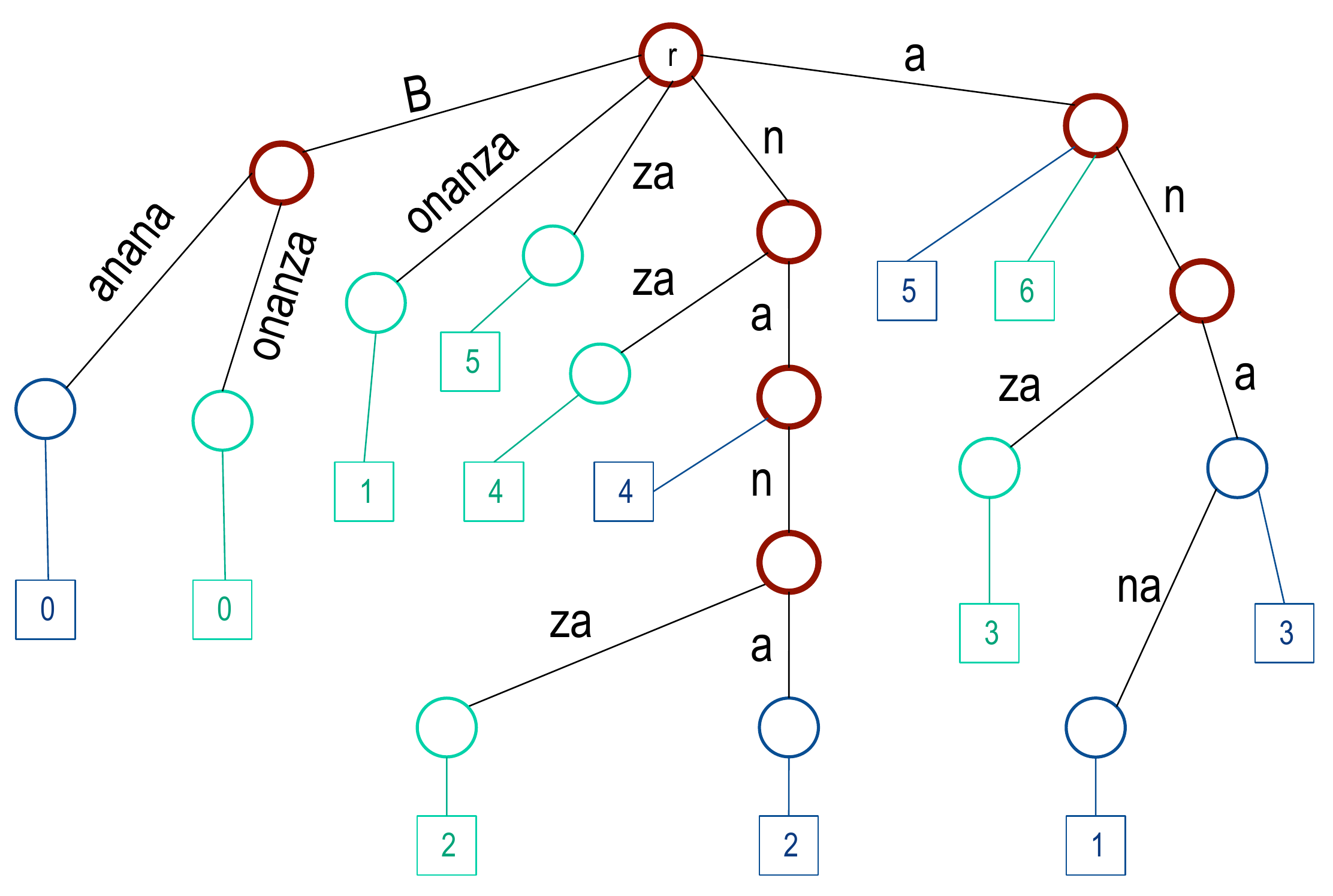}
\caption{Generalized Suffix Tree for {\em Banana} and {\em Bonanza}.}
\label{banabona.fig}
\end{figure}

As an example, consider the generalized suffix tree for the two sequences {\em
  Banana} and {\em Bonanza} as given in Figure~\ref{banabona.fig}. In order to
enhance readability, the leaf nodes for {\em Banana} and {\em Bonanza} are
denoted by blue and green squares, respectively. All fully coloured nodes are
highlighted in red. As we can infer from Figure~\ref{banabona.fig}, the two
sequences contain the common sub-sequences \texttt{B}, \texttt{n},
\texttt{na}, \texttt{nan}, \texttt{a}, and \texttt{an}, respectively.

%%% Image of the suffix tree for 'Banana' and 'Bonanza'

% ------------------------------------------------------------------------ %

\subsection{Multi Sub-words}
\label{multi-subwords.subsec}

The concatenation of the edge labels to all fully coloured branching nodes
give us the {\em values} of the common sub-sequences, but not their respective
{\em positions} (or start indices) in the sequences of a GST. The start
indices are defined by the labels of all leaf nodes in the corresponding
sub-trees of each of the fully coloured branching nodes.
%
%%%% Not sure whether this is necessary, but let's see...
%
For example, the bottom-most fully coloured branching node in Figure
\ref{banabona.fig} represents the common sub-sequence {\tt nan}. Its subtree
contains leafnodes for index $2$ for both sequences, hence {\tt nan} starts
at index $2$ in {\em Banana} and {\em Bonanza}, respectively.

If there is more than one leaf node for a specific colour (or sequence), then
the corresponding sub-sequence appears {\em multiple times} in this sequence,
and multiple combinations can be found across all sequences. For example, the
sub-sequence \texttt{a} appears three times in {\em Banana} and twice in {\em
  Bonanza}, hence a total of $6$ combinations for \texttt{a} can be
found. More generally, if we have a common sub-sequence $\alpha$ that appears
$m$ times each in $n$ sequences, there is a total of $m^n$ possible
combinations. Therefore, MSA approaches (such as the one proposed in
\cite{hoehl:02a}) that rely on generating all possible combinations during an
alignment process, are unlikely to scale to a large number of sequences.

In order to avoid enumerating all possible combinations for a given common
sub-sequence, we introduce the notion of a {\em multi sub-word} (or MSW).
A multi sub-word is a combination of the value of a common sub-sequence,
together with a set of occurrences thereof across all sequences. We use the
notation

\smallskip
\centerline{{\MSW{a}{0@1 0@3 0@5 1@3 1@6}}}

\noindent for multi sub-words, in this case, to indicate that sub-sequence
\texttt{a} appears at indices $1$, $3$ and $5$ in sequence $0$ ({\ie} {\em
  Banana}) and at indices $3$ and $6$ in sequence $1$ ({\ie} {\em Bonanza}),
respectively.

%% file: algorithm.tex
\section{Multiple Sequence Alignment}
\label{msa.sec}

Multiple sequence alignment (MSA) was first used to align three or more
biological sequences to reveal their structural commonalities
\cite{durbin:98a}. In contrast to pair-wise sequence alignment algorithms such
as Needleman-Wunsch \cite{needleman:70a} or Smith-Waterman \cite{smith:81a}
that have a quadratic time and space complexity, creating an optimal alignment
for multiple sequences is an NP-complete problem \cite{wang:94a}. Therefore, a
number of heuristic techniques were proposed, with {\em ClustalW}
\cite{thompson:94a} probably being the most prominent one.

A brief overview of the {\em ClustalW} algorithm is as follows:

\begin{enumerate}
\item All $n (n\!-\!1) / 2$ possible pairs of sequences are aligned by using a
  standard Needleman-Wunsch algorithm \cite{needleman:70a} in order to
  calculate their {\em similarity}, resulting in an $n \times n$ similarity
  matrix. 

\item A \emph{guide tree} is constructed from the similarity matrix by
  applying a neighbour-joining clustering algorithm \cite{saitou:87a}.

\item The tree is then used to guide a progressive pair-wise alignment
  process by traversing the tree from the leave nodes to the root.
\end{enumerate}

\noindent ClustalW has a polynomic algorithmic complexity, but the first step
alone is $\mathcal{O}(n^2 m^2)$ with $m$ being the average sequence length.

%%% NOTE: this may be too much for the paper as we are slowly running out
%%%       of space...
%% \subsection{Graph-based Algorithm}
\label{graphs.subsec}

\smallskip

In order to better exploit similarities between sequences, a number of
GST-based MSA algorithms have been proposed ({\cf}
\cite{bieganski:94a,myers:95a,delcher:99a,hoehl:02a} just to name a few).
H{\"o}hl {\etal} \cite{hoehl:02a}, for example, proposed a MSA algorithm that
(i) creates a GST for the sequences to be aligned, (ii) identifies all nodes
in the GST that adhere to a property equivalent to the notion of fully
coloured, and then (iii) generates all possible combinations of sub-sequences
that are {\em maximal}. A sub-sequence is maximal if it cannot be extended
either to the left or right without having at least one character that is not
common across all sequences. The collection of maximal common sub-sequences
forms the basis of a directed, edge-weighted graph $(V,E)$ with the maximal
common sub-sequences as vertices. A weighted, directed edge is added between
$s_1$ and $s_2$ if and only if $s_2$ is ``to the right'' of $s_1$ in all
sequences. The weight of this edge is the length of $s_1$. The best possible
alignment for all sequences is then defined by a chain through the graph with
maximal weight-length \cite{hoehl:02a}.

Consider the application of this algorithm to the two sequences
\texttt{ADCxzDCxBAx} and \texttt{DCxAzDCxpxBA}. This leads to the following,
non-optimal alignment:

\newcommand{\graphAlign}{
\begin{minipage}[b]{0.45\linewidth}
\texttt{\noindent
ADCx{\agap}zDCxBAx{\agap}{\agap} \newline
{\agap}DCxAzDCxp{\agap}xBA
}
\end{minipage}
}

\noindent {\centerline {\graphAlign}}

%%% Note that this may need a better explanation...
There is a total of $24$ different common sub-sequences, only $12$ thereof are
both right- and left-maximal. However, as \texttt{zDCx} and \texttt{xBA}
overlap in the first sequence (they do not overlap in the second sequence),
% and there is no trimming,
\texttt{xBA} is {\em not} to the right of \texttt{zDCx}, only the last
\texttt{x} in both sequences are. Hence, as \texttt{BA} is not a maximal
sub-sequence and the algorithm does not allow for trimming of partially
overlapping sub-sequences, \texttt{BA} is not aligned, only \texttt{x}
is. Therefore, besides the problem of a possible very large number of all
common sub-sequence combinations ({\cf} Section \ref{multi-subwords.subsec}),
this algorithm does not cater well for partial overlaps and may generate
sub-optimal alignments.

\smallskip

Common to all MSA algorithms we investigated is that they were designed to
align {\em few}, but possibly long sequences. This contrasts with our problem
of creating prototypes for clusters that contain {\em many}, but possibly
short(er) sequences. This mismatch motivated the new MSA algorithm presented
in the following section.

%%%%%%%%%%%%%%%%%%%%%%%%%%%%%%%%%%%%%%%%%%%%%%%%%%%%%%%%%%%%%%%%%%%%%%%%%%%%

\section{Mandile-Schneider Algorithm}
\label{msalgo.sec}

In order to address the scalability issues of existing MSA algorithms, we
defined a new approach that (i) exploits the benefits of using a GST to
identify common sub-sequences and (ii) uses {\em divide-and-conquer} to
align segments that progressively become smaller. In line with other alignment
algorithms (such as Needleman-Wunsch \cite{needleman:70a} and Smith-Waterman
\cite{smith:81a}), we call this algorithm the {\MS} algorithm. The guiding
principles behind the algorithm are that we want to (i) exploit commonalities
between sequence as early as possible, (ii) avoid the enumeration of all
possible combinations of a common sub-sequence (either entirely or as much as
possible) as this can become very computationally intensive, and (iii) give
precedence to long common sub-sequences over shorter ones when applying the
divide-and-conquer principle. Most importantly, though, the approach should be
fit for purpose for our specific needs, that is, the alignment of many, but
short sequences. The main steps of the new algorithm are outlined in
Figure~\ref{msalgo.tab}.

\begin{figure}[t]
\renewcommand{\labelenumi}{\hspace{-1.2cm}[\arabic{enumi}]}
\begin{framed}
\begin{sf}
\begin{enumerate}
\item \label{gst.step} Create a generalized suffix tree for the sequences to
  be aligned.

\item \label{msw.step} Identify all {\em fully-coloured branching nodes} and
  create a corresponding collection of multi sub-words.\\[-1mm]

\item \label{best-mws.step} Choose the ``best'' multi sub-word {\sc msw} from
  this collection.

\item \label{best-csw.step} Choose the ``best'' common sub-sequence {\sc csw}
  from the multi sub-word {\sc msw}. {\sc csw} will act as {\em anchor} for the
  first alignment.

\item \label{overlap.step} For each multi sub-word in the collection, remove
  any sub-sequences that {\em fully overlap} with {\sc csw}.

\item \label{trimm.step} For each multi sub-word in the collection, trim any
  sub-sequences that {\em partially overlap} with {\sc csw}. Due to the
  trimming, this step can either add new multi-subwords or add sub-sequence
  information to existing multi sub-words.

\item \label{split.step} Split the collection of multi sub-words into two
  collections: one that is {\em entirely to the left} and one that is {\em
    entirely to the right} of the anchor {\sc csw}.\\[-1mm]

\item Recursively apply steps \ref{best-mws.step} and \ref{split.step} to
  both, the left and right collections. Terminate the recursion at Step
  \ref{best-mws.step} if the collection of multi sub-words is empty.

\end{enumerate}
\end{sf}
\end{framed}
%% {\figureline}
\caption{The main steps of the {\MS} algorithm.}
\label{msalgo.tab}
\end{figure}

The result of the algorithm is a list of common sub-sequences that defines all
anchors of the alignment. A gap between two consecutive sub-sequences is an
area where no alignment was found.

\smallskip

The reader may note that the identification of the ``best'' multi sub-word and
common sub-sequence in steps \ref{best-mws.step} and \ref{best-csw.step},
respectively, is left open. These are two {\em variation points} in the
algorithm where different strategies can be adapted to identify the ``best''
anchor. In line with the guiding principles mentioned earlier, we use a
{\bigleftmost} strategy for the majority of this work, that is, the {\em
  left-most} common sub-sequence in the {\em largest} multi sub-word is chosen
as the anchor for each alignment step.{\footnote{The current implementation
    chooses the {\em first} multi sub-word if more than one largest is found.}}

% ------------------------------------------------------------------------ %

%% \subsection{Example}
\smallskip

Let's illustrate the main steps of the {\MS} algorithm on the two sequences
\texttt{ADCxzDCxBAx} and \texttt{DCxAzDCxpxBA} used in the discussion of the
graph search-based approach by H{\"o}hl {\etal} \cite{hoehl:02a}:

\renewcommand{\labelitemi}{\hspace{-0.5cm} $\bullet$}
\begin{itemize}

\item {\bf Step \ref{gst.step} and \ref{msw.step}:} the GST of the
  two sequences has $7$ fully coloured branching nodes, leading to the
  following collection of multi sub-words:

%   \noindent
%   \begin{footnotesize}
%   \begin{tabular}{lcl}
%     $[$ A  & -- & \{ 0@0 0@9 1@3 1@11 \} $]$ \\
%     $[$ BA & -- & \{ 0@8 0@10 \} $]$ \\
%   \end{tabular}
%   \end{footnotesize}

  \begin{footnotesize}
  {\MSW{\texttt{A}}{0@0 0@9 1@3 1@11}} \\
  {\MSW{\texttt{BA}}{0@8 0@10}} \\
  {\MSW{\tt Cx}{0@2 0@6 1@1 1@6}} \\
  {\MSW{\tt DCx}{0@1 0@5 1@0 1@5}} \\
  {\MSW{\tt x}{0@3 0@7 0@10 1@2 1@7 1@9}} \\
  {\MSW{\tt xBA}{0@7 1@9}} \\
  {\MSW{\tt zDCx}{0@4 1@4}}
  \end{footnotesize}

\item {\bf Step \ref{best-mws.step} and \ref{best-csw.step}:} the longest
  multi sub-word is \texttt{zDCx}, the left-most common sub-sequence thereof
  starts at index $4$ in both words. This becomes the first anchor.

\item {\bf Step \ref{overlap.step}:} the removal of any overlaps with
  \texttt{zDCx} at index $4$ in both words reduces the collection of multi
  sub-words to the following:

  \begin{footnotesize}
  {\MSW{\texttt{A}}{0@0 0@9 1@3 1@11}} \\
  {\MSW{\texttt{BA}}{0@8 0@10}} \\
  {\MSW{\tt Cx}{0@2 1@1}} \\
  {\MSW{\tt DCx}{0@1 1@0}} \\
  {\MSW{\tt x}{0@3 0@10 1@2 1@9}} \\
  {\MSW{\tt xBA}{0@7 1@9}} \\
  \end{footnotesize}

  \vspace{-0.2cm} \texttt{zDCx} overlaps with itself, hence is fully removed,
  and one occurrence each of \texttt{DCx} and \texttt{x} is removed.

\item {\bf Step \ref{trimm.step}:} the only partially overlapping
  multi-subword is {\tt xBA} -- it is trimmed back to \texttt{BA}. Since we
  already have a multi sub-word for {\tt BA} at indices $8$ and $10$,
  respectively, the trimmed multi sub-word does not further contribute to the
  collection of multi sub-words.

\item {\bf Step \ref{split.step}:} splitting the multi sub-words into
  what is entirely to the left and right of \texttt{zDCx}, respectively,
  results in the following two collections of multi sub-words:

  \medskip
  \begin{footnotesize}
  \begin{tabular}{ll}
  {\em {\small Left:}} & {\em {\small Right:}} \\
  {\MSW{\texttt{A}}{0@0 1@3}}\ \ \  & {\MSW{\texttt{A}}{0@9 1@11}} \\
  {\MSW{\tt Cx}{0@2 1@1}}           & {\MSW{\texttt{BA}}{0@8 0@10}} \\
  {\MSW{\tt x}{0@3 1@2}}            & {\MSW{\tt x}{0@10 1@9}} \\
  {\MSW{\tt DCx}{0@1 1@0}}\ \ \ \ \ & \\
  \end{tabular}
  \end{footnotesize}

  \medskip
  Please note that {\tt A} and {\tt x} occur in both the left and right
  collections as there is one occurrence each on either side of the anchor
  {\tt zDCx}.
  
\item {\bf Recursion on {\em left}:} {\tt DCx} is the longest multi sub-word
  in {\em left}, the left-most common sub-sequence thereof starts at indices
  $1$ and $0$, respectively. The removal of overlaps only leaves {\tt A}, but
  since {\tt A} {\em crosses over} the anchor {\tt DCx} (the occurrence of {\tt
    A} in the first sequence is left of the anchor, but to the right in the
  second sequence), the resulting left and right collections in Step
  \ref{split.step} are empty, and the recursion for both terminates in Step
  \ref{best-mws.step}.

\item {\bf Recursion on {\em right}:} {\tt BA} is the longest multi sub-word
  in {\em right}, the left-most common sub-sequence thereof starts at indices
  $8$ and $10$, respectively. Similar to the recursion on {\em left}, the
  removal of overlaps leaves one multi sub-word only ({\ie} {\tt x}) that
  crosses over the anchor, hence the resulting left and right collections in
  Step \ref{split.step} are empty, and the recursion terminates as well.

\item {\bf Result:} we get three anchors {\tt DCx}, {\tt zDCx}, and {\tt BA}
  that define the following, optimal alignment:

  \smallskip
  {
  \texttt{\noindent
  \centerline{ADCx{\agap}zDCx{\agap}{\agap}BAx\ \ \ \ } \newline
  \centerline{{\agap}DCxAzDCxpxBA{\agap}\ \ \ \ }}}

%  \noindent {\centerline {\msAlign1}}

\end{itemize}

% ------------------------------------------------------------------------ %

%%% \subsection{Implementation}
\smallskip

We implemented the {\MS} algorithm as a set of Java classes and also
integrated a number of packages from
BioJava{\footnote{\label{biojava.foot}{http://www.biojava.org/}}} needed for
the evaluation presented in the next section. During the implementation, we
applied a number of optimizations in order to improve both, the run-time
performance as well as the memory footprint. However, there is room for
further optimizations. A packaged version of the software is available from
{\small {\sf {http://quoll.ict.swin.edu.au/doc/ms/}}}.

%%%%%%%%%%%%%%%%%%%%%%%%%%%%%%%%%%%%%%%%%%%%%%%%%%%%%%%%%%%%%%%%%%%%%%%%%%%%

% 1. Overlaps are removed
% 2. The MSWMap is split based on the previously chosen CSW
% 3. For each MSW the following is done:
%     a. Each partially overlapping SubWordInfo instance is added into a map with Key: overlap amount and Value: set of SubWordInfos
%     b. This map is iterated over, constructing a new MSW for each entry with Trimmed Value and appropriate starting positions.
%     c. All partial overlaps are removed as they are invalid
% 4. All non overlapping and all newly constructed MSWs are put into 2 new maps
% (left and right)

%%%%%%%%%%%%%%%%%%%%%%%%%%%%%%%%%%%%%%%%%%%%%%%%%%%%%%%%%%%%%%%%%%%%%%%%%%%%

%%%% Backups %%%%

% In this section, we will discuss existing approaches on how the common
% sub-sequences identified in a generalized suffix tree can be used to create a
% multi-sequence alignment, followed by the description of our own new algorithm.
% For the rest of this section, we will use the two sequences
% \texttt{ADCxzDCxBAx} and \texttt{DCxAzDCxpxBA} as a running example for
% illustration purposes.

%%%%%%%%%%%%%%%%%%%%%%%%%%%%%%%%%%%%%%%%%%%%%%%%%%%%%%%%%%%%%%%%%%%%%%%%%%%%

%% file: evaluation.tex
% Document Type: LaTeX
% Master File: evaluation.tex
% Author: Jean-Guy Schneider, Peter Mandile, Steve Versteeg
% Last Modified: Fri Aug 28 2015, 15:29

% Evaluation section for the ASWEC 2015 submissions

%%%%%%%%%%%%%%%%%%%%%%%%%%%%%%%%%%%%%%%%%%%%%%%%%%%%%%%%%%%%%%%%%%%%%%%%%%%%

\section{Evaluation}
\label{eval.sec}

In this section, we present the experiments that we conducted to evaluate the
accuracy and effectiveness of the novel MSA approach presented in the previous
section and discuss the results of our experiments. More specifically, we
introduce our experimental setup as well as our evaluation criteria in
Section~\ref{setup.subsec}. In Section \ref{evalres.subsec}, we present and
discuss the results of our evaluation, especially from a resource usage and
accuracy perspective. We discuss limitations of our current approach and
identify possible areas of future improvements in Section
\ref{disclimit.subsec}. Finally, we present an initial evaluation of the new
MSA algorithm on biological data in Section \ref{biodata.subsec}.

% ------------------------------------------------------------------------ %

\subsection{Experimental Setup}
\label{setup.subsec}

Although one of the main aims of our work is to generate regular expressions
to summarize message clusters for unknown or ill-specified protocols, for
evaluation purposes, we used three protocols where the precise message
structures for the various operation types are known: the Lightweight
Directory Access Protocol (LDAP) \cite{sermersheim:06a}, the Simple Object
Access Protocol (SOAP) \cite{box:00a}, and IBM Information Management System
(IMS) \cite{ims:2000} which is widely used on mainframe computers.

The recorded LDAP messages used for this evaluation consisted of the core LDAP
operations {\sc add}, {\sc search}, and {\sc modify} applied to CA Technologies' {\em
  DemoCorp} sample directory \cite{ca:democorp}. We used a decoder to
translate the corresponding messages into a text format, but kept a subset of
the {\sc search} operations in their binary format.

The SOAP messages used for evaluation purposes were generated based on a
recording of a banking example using the LISA tool \cite{michelsen:11a}. The
protocol consists of $7$ different request types, each with a varying number
of parameters, encoding ``typical'' transactions one would expect from a
banking service. We chose to use the messages from the {\sc withdrawMoney}
operation for this evaluation.

The IMS messages used were recorded interactions between a client tool and an
IMS phone book demonstration server. IMS messages consist of some header data
followed by the record structure encoded as {\em fixed width fields}. The
demonstration server supported four different transaction types including:
{\sc ADD}, {\sc UPDATE}, {\sc DISPLAY} and {\sc DELETE}. In addition the
recording included IMS acknowledgement messages. For the purpose of the
evaluation, we used the messages of the {\sc UPDATE} transaction type. 

A summary of the protocols, operations, number of messages as well as the
median message length (measured in number of characters) is given in
Table~\ref{case-studies.tab}. The raw input data used for the evaluation is
available at {\small {\sf {http://quoll.ict.swin.edu.au/doc/ms/}}}.

%%% The link does not seem to work :-(
%% {\small {\sf {http://quoll.ict.swin.edu.au/doc/message\_traces.html}}}.

%%% Table to summarize the relvant protocols and clusters, respectively
\begin{table}[t]
\centerline{
\begin{tabular}[c]{|l|l|c|c|c|}
\hline
Protocol & Operation & Type & No. Msgs. & Median Length \\
\hline
LDAP & Add      & text   & 851 & 378 \\ 
LDAP & Modify   & text   & 334 & 239 \\
LDAP & Search   & text   & 621 & 313 \\
SOAP & Withdraw & text   & 160 & 243 \\
LDAP & Search   & binary & 606 &  73 \\
IMS  & Update   & binary & 200 & 231 \\
\hline
\end{tabular}
}
\caption{Protocols and operations used for evaluation.}
\label{case-studies.tab}
\end{table}

\smallskip

For the purpose of the evaluation, we were interested in the (i) resource
usage of the new approach and (ii) the quality and accuracy of the generated
alignments. In order to compare the obtained results, we used the ClustalW
\cite{thompson:94a} implementation of BioJava 4.0{\footnoteref{biojava.foot}}
as a comparison benchmark.

Resource usage was measured in both {\em total computation time} for an
alignment (excluding the time to read the input and analyzing the generated
alignment) as well as the {\em memory usage} as reported by the Java Virtual
Machine.{\footnote{This was measured by subtracting the free memory from the
    total memory directly after the generation of the alignment.}} The {\em
  quality} of each generated alignment is measured using an {\em edit
  distance} \cite{ristad:98a} as well as the total number of overlapping ({\ie}
aligned) characters. With regards to accuracy, for each of the operation types
of the six protocols/operations used, we identified a number of sub-sequences
(representing specific structural information of the corresponding operation
types) and checked whether these sub-sequences were aligned across all
messages.

\smallskip

For each of the six input sets used for evaluation, the input files with the
corresponding messages were {\em randomly} ordered in order to avoid any bias
from the corresponding recordings. Therefore, in order to create an alignment
for $n$ messages, we simply used the $n$ first lines in the corresponding file
as input.

For each protocol/operation, we generated alignments for 2, 5, 10, 15, 20, 25,
50, 100, and 200 messages as well as the maximal available number of messages
in order to illustrate the scalability of both, the proposed approach as well
as the comparison benchmark. Each alignment computation was repeated {\em 25
  times} in order to mitigate against any ``random'' effects from other
processes running on the computing platform. 

All experiments were run on an iMac with a 2.7 GHz Intel Core 5 CPU, 8GM of
RAM, running Mac OS X 10.8.5 and Java 1.8.

% ------------------------------------------------------------------------ %

\subsection{Evaluation Results}
\label{evalres.subsec}

The results of our evaluation experiments are summarized in
Table~\ref{results.tab}: for each combination of operation type/number of
messages to align it lists: the median total time taken to generate an
alignment (in milli-seconds), the median memory usage (in Megabytes), the edit
distance of the resulting alignment, and the number of overlapping characters
({\ie} the sum of the lengths of all common sub-sequences in the alignment).
For the proposed {\MS} approach, we also list the number of multi sub-words
that were extracted from the generalized suffix tree (not applicable to
ClustalW). The final table column, titled `Speed-Up', shows the performance
increase of {\MS} compared to ClustalW.

%%% One BIG table with all the relevant results
\input{results}

% - Timing, speedup
%    - quadratic (cubic?) time complexity of ClustalW shows...
%    - almost linear time for MS => more experiments to illustrate required...

\subsubsection{Resource Usage}

Table~\ref{results.tab} illustrates the differences in time complexity for the
proposed algorithm compared to ClustalW, respectively: for all data points,
the total time taken for {\MS} is significantly better than the one for
ClustalW. The speed-ups vary from $3.78$ for 2 sequences in LDAP Search to
$55.63$ for 851 sequences in LDAP Add.

A brief statistical analysis demonstrated a strong quadratic correlation ($R^2
> 0.99$) between the number of sequences and the total alignment time for
ClustalW for all of the six protocols/operations. This is not surprising as
for $n$ sequences, ClustalW needs to compute the similarities of all $n*(n\!
-\!1)$ possible pairs in order to build the guide tree. The corresponding
(quadratic) coefficients varied from $0.097$ for the binary LDAP Search to
$2.50$ for LDAP Add. However, even for binary LDAP Search (where the
individual messages are short compared to the other five case examples), the
quadratic correlation was much stronger than a corresponding linear
relationship.
On the other hand, our proposed algorithm exhibits a linear correlation for
the number of sequences/computation time relationship for LDAP Modify, SOAP,
binary LDAP Search and IMS, respectively, whereas the other two
protocols/operations showed a quadratic correlation (but with coefficients
around $0.03$ for the quadratic term).

%%% Shall I mention something about linear time to create the GST here,
%%% hence 'the rest' 

We can conclude that the proposed algorithm is faster than the ClustalW
implementation of BioJava 4.0, and the more sequences need to be aligned, the
greater the speedup. To some extent this is not surprising given the
algorithmic complexity of ClustalW. However, we will need to run further
experiments on different kind of protocols and/or operations and perform both,
a detailed statistical analysis as well as an algorithmic complexity analysis
to get further evidence of a possible linear time complexity of the proposed
algorithm.

%%%% The underlying data for the discussion above... %%%
% ldapAdd: .0389619  vs 2.498449 , bio higher constants for 
% very high R-squared (0.99)
%
% ldapSearch:  .0284735 vs 1.591302
%
% ldapModify: linear for MS vs. 1.060026 (but linear not a good fit for bio)
%
% SOAP: linear for MS, 1.056575 (but linear not a good fit for bio)
%
% LDAP bin: linear for MS, .0961957 (but linear not a good fit for bio)
%
% IMS: linear for MS, .6308246 (but linear not a good fit for bio)
%%%%

% - Memory requirements -> why BioJava so inconsistent
%    - clearly not consistent pattern, but due to 'short' nature of messages,
%      comparable memory requirements. Garbage collection
%      On average, MS probably better. For longer messages, see biodata

\smallskip

\begin{figure}[t]
\centering
\includegraphics[width=3.4in]{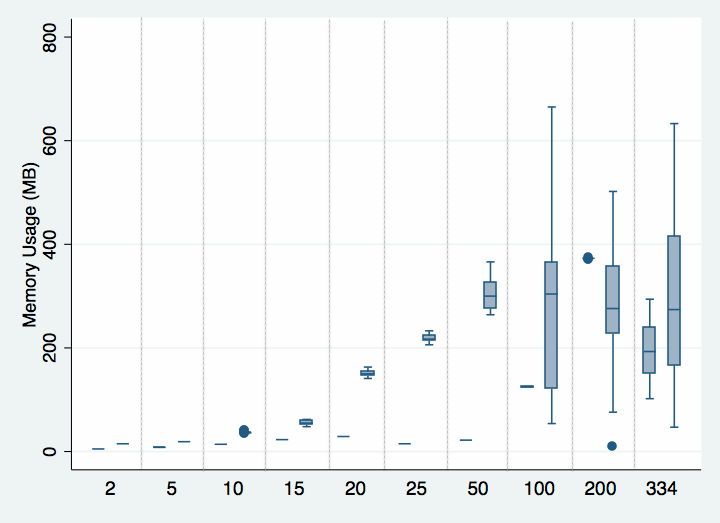}
\caption{Memory usage for LDAP Modify: {\MS} on the left, ClustalW on the
  right. Number of sequences to be aligned on the x-axis.}
\label{memldapmod.fig}
\end{figure}

The results for the memory usage are unfortunately not as consistent. Consider
the results for all sequence combinations for LDAP modify as given in
Figure~\ref{memldapmod.fig}.

For ClustalW, once the similarity between two sequences is established, the
matrix required to compute the similarity is not needed any more and hence can
be garbage collected. Therefore, the larger the number of sequences, the more
likely the garbage collector will kick in, but as Figure~\ref{memldapmod.fig}
illustrates, this does not happen in a consistent manner ({\eg} demonstrated
by the large spread for 100 sequences). Similarly, for the proposed new
approach, once all multi sub-words are identified, the generated GST is not
needed any more and can be garbage collected. We suspect this has happens as
the median memory usage for all $334$ sequences for LDAP Modify is lower than
for $200$ sequences. However, without a more accurate measurement of the
maximal memory usage we cannot really draw too many conclusions from the
experiments.

%%% Graph of the memory usage of LDAP Modify

\subsubsection{Accuracy}

% - Accuracy, Edit Distance, Overlap
%     - comparable edit distance for most messages (except IMS)
%     - not identical, but main message structures are aligned
%     - ClustalW -> sub-alignments, not in MS
%     - nature of the chosen operations -> structure larger than payload

To the best of our knowledge, the generated alignments of both approaches do
separate message structure information from payload, precisely as was
required. This was verified by a combination of manual inspection and
hand-crafted regular expressions that were run over the generated alignments.
With the exception of IMS Update (which will be discussed in more detail in
the next section), the edit distances of {\MS} are no more than 10\% worse
than ClustalW. The differences in edit distance between the two alignment
approaches is mainly due to how the payload is aligned -- structure
information is aligned consistently. Whereas ClustalW minimizes the number of
gaps, {\MS} maximizes the total number of overlapping characters which may
result in larger gaps between aligned sub-sequences, and, consequently, a
larger edit distance. Initial results using the generated alignments to create
regular expressions as cluster prototypes (the main aim of our investigation)
indicate that the alignments are accurate enough to do so, but a further, more
detailed analysis is required to substantiate this claim. This is the topic of
an ongoing investigation.

\smallskip

%%% Discussion about total number of common sub-words -> shows that the
%%% graph-based approach just does not scale!
An interesting observation of our experiments is that the total number of {\em
  multi sub-words} generated by the corresponding generalized suffix tree is
quite low. In case of LDAP Add, for example, even the GST for all 851 messages
does not define more than 447 multi sub-words ({\cf} Table~\ref{results.tab}).
On the other hand, the total number of {\em common sub-sequences} contained in
these multi sub-words grow exponentially. For the first two messages, there
are \numprint{26 339} common sub-subsequences, growing to
\numprint{13906721076} ({\ie} more than 13 Billion) for five messages. For 10
messages, the total number of common sub-sequences cannot be represented using
a 64 Bit integer any more ({\ie} it is greater than $10^{18}$), demonstrating
that the graph-based approach introduced by H{\"o}hl {\etal} \cite{hoehl:02a}
does indeed not scale beyond small examples as postulated in
Section~\ref{graphs.subsec}.

% ------------------------------------------------------------------------ %

\subsection{Discussion and Limitations}
\label{disclimit.subsec}

%%% NOTE: if there is some time (we have a bit of space), mention the
%%% structure/payload ration of the text-based operations. The question
%%% then is what we do with binary LDAP as it is the odd one out and
%%% probably needs some discussion...

% -> Need to focus on why IMS does not work...
As illustrated in Table~\ref{results.tab}, there is one protocol/operation
type that performs much worse with regards to the edit distance for the
proposed algorithm compared to ClustalW: IMS Update. In the following, we
briefly discuss one of the main reasons for this significantly worse edit
distance.

\newcommand{\aSpace}{\makebox[\aligncharw][s]{\textvisiblespace}}
\newcommand{\imsPayload}{
    \begin{minipage}[b]{.85\linewidth}
    \texttt{\footnotesize{\noindent
Garufi{\aSpace}{\aSpace}{\aSpace}{\aSpace}Meaghan{\aSpace}{\aSpace}{\aSpace}598{\aSpace}{\aSpace}{\aSpace}{\aSpace}{\aSpace}{\aSpace}{\aSpace}37110 \newline
Hoogland{\aSpace}{\aSpace}Tamar{\aSpace}{\aSpace}{\aSpace}{\aSpace}{\aSpace}575{\aSpace}{\aSpace}{\aSpace}{\aSpace}{\aSpace}{\aSpace}{\aSpace}14895 \newline
Lindall{\aSpace}{\aSpace}{\aSpace}Fabian{\aSpace}{\aSpace}{\aSpace}{\aSpace}371{\aSpace}{\aSpace}{\aSpace}{\aSpace}{\aSpace}{\aSpace}{\aSpace}80126 \newline
Fern{\aSpace}{\aSpace}{\aSpace}{\aSpace}{\aSpace}{\aSpace}Natalie{\aSpace}{\aSpace}{\aSpace}937{\aSpace}{\aSpace}{\aSpace}{\aSpace}{\aSpace}{\aSpace}{\aSpace}82901
}}
\end{minipage}
}

Consider the following extract of the payload information (surname, given
name, area code, phone number) for four messages that each contain a varying
number of white spaces (shown as `\aSpace') in order to ensure a fixed-length 
message encoding:

\noindent {\imsPayload}

\newcommand{\imsMS}{
\begin{minipage}[b]{\linewidth}
\texttt{\footnotesize{\noindent
Garufi**********{\aSpace}{\aSpace}{\aSpace}{\aSpace}Meaghan{\aSpace}{\aSpace}{\aSpace}598**{\aSpace}{\aSpace}{\aSpace}{\aSpace}{\aSpace}{\aSpace}{\aSpace}37**110** \newline
Hoogland{\aSpace}{\aSpace}Tamar*{\aSpace}{\aSpace}{\aSpace}{\aSpace}{\aSpace}575***********{\aSpace}{\aSpace}{\aSpace}{\aSpace}{\aSpace}{\aSpace}{\aSpace}****14895 \newline
Lindall{\aSpace}{\aSpace}{\aSpace}Fabian{\aSpace}{\aSpace}{\aSpace}{\aSpace}371************{\aSpace}{\aSpace}{\aSpace}{\aSpace}{\aSpace}{\aSpace}{\aSpace}80**126** \newline
Fern************{\aSpace}{\aSpace}{\aSpace}{\aSpace}{\aSpace}{\aSpace}Natalie{\aSpace}{\aSpace}{\aSpace}937{\aSpace}{\aSpace}{\aSpace}{\aSpace}{\aSpace}{\aSpace}{\aSpace}82901****
}}
\end{minipage}
}

\smallskip
The {\MS} algorithm generates the following, non-optimal alignment with
an edit distance of 109:

\noindent {\imsMS}

\vspace{-1.8mm}
The longest common sub-sequence is the seven white spaces between the area
code and phone numbers and this sequence is (correctly) aligned first. The
longest sub-sequence to its left are four white spaces: between
\texttt{Garufi} and \texttt{Meaghan} for the first sequence, the first four
white spaces between \texttt{Tamar} and \texttt{575} for the second, between
\texttt{Fabian} and \texttt{371} for the third, and finally the first four
white spaces between \texttt{Fern} and \texttt{Natalie}, respectively. Once
this sub-sequence is aligned, the rest ``in the middle'' does not contain any
common characters any more and cannot be further aligned.

Please also note that all phone numbers contain at least one number
`\texttt{1}' and the left-most occurrences thereof are aligned. One may argue
that such short common sequences should not be aligned and it would be
straightforward to define a threshold to ensure that any alignments shorter
than this threshold are ignored. However, the binary LDAP search has a single
character `\texttt{c}' that indicates the operation type and if we introduce
such a threshold then the \texttt{c}'s are unlikely to be aligned. Therefore,
we decided against introducing such a threshold.

\newcommand{\imsClusalW}{
    \begin{minipage}[b]{.85\linewidth}
    \texttt{\footnotesize{\noindent
Garufi{\aSpace}{\aSpace}{\aSpace}{\aSpace}Meaghan{\aSpace}{\aSpace}{\aSpace}598*{\aSpace}{\aSpace}{\aSpace}{\aSpace}{\aSpace}{\aSpace}{\aSpace}37110 \newline
Hoogland{\aSpace}{\aSpace}Tamar{\aSpace}{\aSpace}{\aSpace}{\aSpace}{\aSpace}575*{\aSpace}{\aSpace}{\aSpace}{\aSpace}{\aSpace}{\aSpace}{\aSpace}14895 \newline
Lindall{\aSpace}{\aSpace}{\aSpace}Fabian*{\aSpace}{\aSpace}{\aSpace}{\aSpace}371{\aSpace}{\aSpace}{\aSpace}{\aSpace}{\aSpace}{\aSpace}{\aSpace}80126 \newline
Fern{\aSpace}{\aSpace}{\aSpace}{\aSpace}{\aSpace}{\aSpace}Natalie{\aSpace}{\aSpace}{\aSpace}937*{\aSpace}{\aSpace}{\aSpace}{\aSpace}{\aSpace}{\aSpace}{\aSpace}82901
}}
\end{minipage}
}

In contrast, ClustalW generates a much better alignment (edit distance of 60),
but even this alignment is not optimal:

\noindent {\imsClusalW}

Aligning the \texttt{1}'s in the phone numbers would introduce new gaps, hence
this is omitted, but it is not clear why a gap between \texttt{Fabian} and the
following white spaces is introduced, but not for \texttt{Tamar}.

% -> Need to mention different strategies for future work
% -> Bias due to structure larger than payload, but not really for binary LDAP!

% ------------------------------------------------------------------------ %

\subsection{Biological Data}
\label{biodata.subsec}

%%% Graph of the alignment comparison
\begin{figure}[t]
\centering
\includegraphics[width=3.4in]{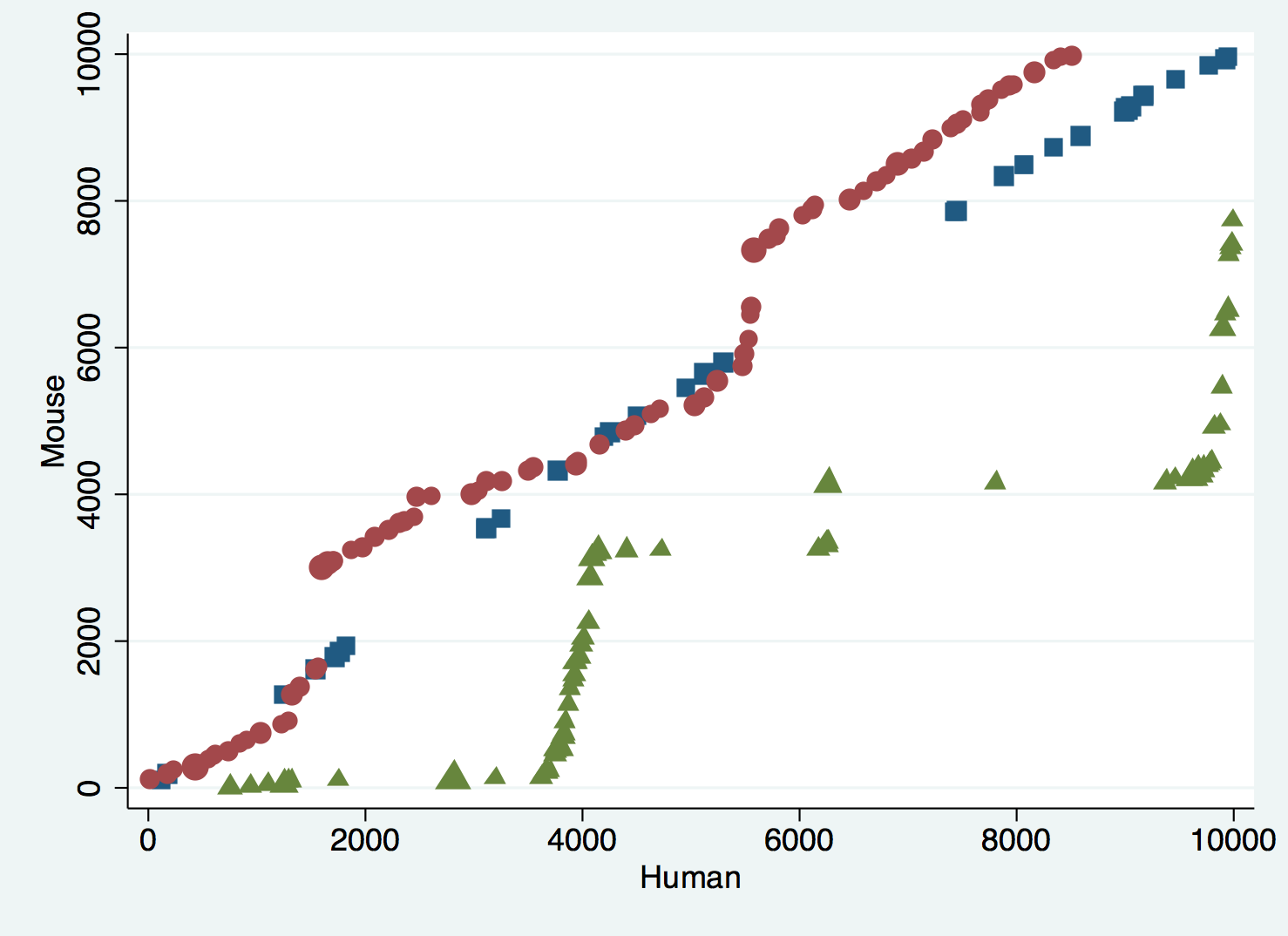}
\caption{Alignments of Human and Mouse genomes.
Blue squares: Needleman-Wunsch.
Green triangles: \MS (biggest-left-most-common-subword).
Red circles: \MS (9 largest multi sub-words).}
\label{humanmouse.fig}
\end{figure}

Aligning two (or more) sequences of nucleotide or amino acid residues to
identify regions of similarity that may be a consequence of functional,
structural, or evolutionary relationships is standard practice in
bioinformatics. Hence, in order to test the accuracy of the new alignment
algorithm with different input data, we chose to align the first \numprint{10
  000} amino acids of both, the human and the mouse genome.{\footnote{Refer to
    \cite{delcher:99a} for specific details about the two genomes.}} As a
benchmark, we used the alignment generated by the ``standard''
Needleman-Wunsch algorithm \cite{needleman:70a}.

Figure~\ref{humanmouse.fig} summarizes the results of this experiment. We only
plot an alignment if an aligned sub-sequence is more than 5 characters ({\ie}
amino acids) long. The size of each symbol relates to the total number of
characters in the corresponding sub-sequence. The alignment of the benchmark
Needleman-Wunsch algorithm (represented as blue squares in
Figure~\ref{humanmouse.fig}) generated only one sub-sequence of 10 characters
with 33 sub-sequences (out of a total of \numprint{3479}) having a length
greater than 5. This indicates that the human and the mouse genomes do not
have a large overlap in the first \numprint{10 000} amino acids.

The {\bigleftmost} strategy does not generate an alignment that is comparable
to the one produced by the benchmark Needleman-Wunsch -- it is represented by
green triangles in Figure~\ref{humanmouse.fig}. The largest alignment is 20
characters long and there are a total of 62 sub-sequences (out of 244) with a
size greater than 5. However, the largest alignment is found at index
\numprint{2819} in the human genome, but \numprint{98} in the mouse, a
significant difference between the two. An algorithm that favours size (of
sub-sequences) over ``location'' is probably not a good fit to align the
first  \numprint{10 000} amino acids of the genomes of human and mouse.

Therefore, we devised another strategy to identify the anchor at each step of
the alignment. Instead of getting the left-most common sub-sequence of the
largest multi sub-word, we identify the $n$ largest multi sub-words, generate
all common sub-sequences thereof, and choose the common sub-sequence with the
smallest variance in the relative starting indices (normalized by size) as the
new anchor. Figure~\ref{humanmouse.fig} includes the alignment generated by
this strategy (represented by red circles) with $n = 9$. The largest alignment
is 15 characters long and there are a total of 81 common sub-sequences (out of
\numprint{1645}) with a size greater than 5.

Clearly, the revised strategy seems to be a much better fit than the original
{\bigleftmost} strategy. It also shows the flexibility of the proposed approach
as new anchor identification strategies can be defined that better fit the
specific nature of a given alignment problem at hand.

%%%%%%%%%%%%%%%%%%%%%%%%%%%%%%%%%%%%%%%%%%%%%%%%%%%%%%%%%%%%%%%%%%%%%%%%%%%%

%%% To discuss here:
%  - example scenarious (which protocols to use)
%  - environment of evaluation and 'set-up'
%  - results of the various 'runs'
%  - discussion of results, possibly limitations

%%%%%%%%%%%%%%%%%%%%%%%%%%%%%%%%%%%%%%%%%%%%%%%%%%%%%%%%%%%%%%%%%%%%%%%%%%%%

%% file: results.tex
% Document Type: LaTeX
% Master File: results.tex
% Author: Jean-Guy Schneider
% Last Modified: Tue Jun 9 2015, 21:03

% Summary table of all experiments

%%%%%%%%%%%%%%%%%%%%%%%%%%%%%%%%%%%%%%%%%%%%%%%%%%%%%%%%%%%%%%%%%%%%%%%%%%%%

\begin{table*}[t]
\centerline{
\begin{tabular}[c]{|l|r||r|r|r|r|c||r|r|r|r||c|}
\hline
   & & \multicolumn{5}{c||}{\small{Mandile-Schneider}} & \multicolumn{4}{c||}{\small{ClustalW}} &  \\[0.2cm]
Protocol & \#Seq. & Time(ms) & Mem.(MB) & Edit-Dist. & \#Chars & \#MSWs & Time(ms) & Mem.(MB) & Edit-Dist. & \#Chars & Speed-Up \\
\hline\hline
 & 2 & 82 & 9 & 119 & 290 & 376 & 309 & 23 & 107 & 288 & 3.77 \\
 & 5 & 109 & 15 & 579 & 267 & 368 & 719 & 14 & 481 & 256 & 6.60 \\
 & 10 & 158 & 27 & 1279 & 263 & 389 & 1324 & 40 & 1238 & 257 & 8.38 \\
 & 15 & 212 & 20 & 2089 & 263 & 391 & 2041 & 121 & 2053 & 253 & 9.63 \\
LDAP Add & 20 & 266 & 11 & 3019 & 263 & 397 & 2959 & 69 & 2870 & 254 & 11.12 \\
 & 25 & 310 & 42 & 3945 & 263 & 401 & 4198 & 304 & 3662 & 254 & 13.54 \\
 & 50 & 523 & 85 & 8635 & 263 & 407 & 10656 & 218 & 8402 & 252 & 20.37 \\
 & 100 & 1069 & 190 & 18475 & 263 & 416 & 33103 & 260 & 17483 & 252 & 30.97 \\
 & 200 & 2674 & 206 & 37511 & 263 & 430 & 113122 & 192 & 36738 & 250 & 42.30 \\
 & 500 & 12925 & 416 & 97875 & 263 & 441 & 667836 & 260 & 93450 & 246 & 51.67 \\
 & 851 & 33701 & 761 & 166967 & 263 & 447 & 1874737 & 316 & 159297 & 246 & 55.63 \\
\hline
 & 2 & 60 & 5 & 158 & 161 & 249 & 222 & 15 & 147 & 164 & 3.70 \\
 & 5 & 72 & 8 & 849 & 150 & 230 & 448 & 19 & 614 & 147 & 6.22 \\
 & 10 & 118 & 14 & 2109 & 133 & 174 & 792 & 36 & 1497 & 126 & 6.71 \\
 & 15 & 145 & 23 & 2602 & 132 & 178 & 1146 & 55 & 2448 & 121 & 7.90 \\
LDAP Modify & 20 & 169 & 29 & 3579 & 132 & 178 & 1545 & 150 & 3053 & 110 & 9.14 \\
 & 25 & 207 & 15 & 4546 & 132 & 180 & 2222 & 217 & 3858 & 110 & 10.73 \\
 & 50 & 349 & 22 & 9683 & 132 & 189 & 5376 & 300 & 8394 & 106 & 15.40 \\
 & 100 & 608 & 125 & 20292 & 132 & 199 & 14946 & 304 & 18566 & 109 & 24.58 \\
 & 200 & 1584 & 373 & 43660 & 132 & 208 & 50269 & 276 & 41044 & 109 & 31.74 \\
 & 334 & 2786 & 193 & 73098 & 132 & 214 & 131497 & 274 & 64415 & 112 & 47.20 \\
\hline
 & 2 & 69 & 6 & 36 & 284 & 392 & 261 & 18 & 29 & 282 & 3.78 \\
 & 5 & 95 & 11 & 157 & 283 & 404 & 502 & 11 & 121 & 282 & 5.28 \\
 & 10 & 140 & 23 & 649 & 274 & 394 & 825 & 34 & 563 & 272 & 5.89 \\
 & 15 & 187 & 12 & 1011 & 272 & 406 & 1304 & 59 & 881 & 268 & 6.97 \\
 & 20 & 218 & 30 & 1361 & 272 & 411 & 1819 & 173 & 1185 & 268 & 8.34 \\
LDAP Search & 25 & 266 & 24 & 1893 & 272 & 423 & 2560 & 44 & 1678 & 268 & 9.62 \\
 & 50 & 468 & 42 & 3808 & 272 & 434 & 6447 & 115 & 3384 & 269 & 13.78 \\
 & 100 & 879 & 44 & 7792 & 272 & 449 & 19952 & 280 & 6893 & 268 & 22.70 \\
 & 200 & 2114 & 181 & 17357 & 272 & 465 & 71828 & 322 & 15602 & 268 & 33.98 \\
 & 500 & 9819 & 490 & 45473 & 272 & 487 & 414885 & 341 & 41007 & 268 & 42.25 \\
 & 621 & 14018 & 536 & 56537 & 272 & 490 & 636085 & 354 & 50982 & 268 & 45.38 \\
\hline
 & 2 & 57 & 5 & 11 & 234 & 349 & 224 & 14 & 10 & 233 & 3.93 \\
 & 5 & 71 & 7 & 40 & 230 & 336 & 398 & 18 & 40 & 230 & 5.61 \\
 & 10 & 97 & 11 & 93 & 230 & 334 & 591 & 61 & 93 & 230 & 6.09 \\
 & 15 & 129 & 16 & 145 & 230 & 334 & 922 & 94 & 145 & 230 & 7.15 \\
SOAP Withdraw & 20 & 136 & 22 & 228 & 230 & 334 & 1240 & 67 & 228 & 230 & 9.12 \\
 & 25 & 158 & 29 & 286 & 230 & 334 & 1663 & 95 & 286 & 230 & 10.53 \\
 & 50 & 265 & 21 & 685 & 230 & 334 & 4373 & 568 & 598 & 229 & 16.50 \\
 & 100 & 469 & 87 & 1410 & 230 & 334 & 13380 & 295 & 1226 & 229 & 28.53 \\
 & 160 & 732 & 115 & 2411 & 230 & 335 & 31449 & 229 & 2116 & 229 & 42.96 \\
\hline
\hline
 & 2 & 35 & 3 & 36 & 36 & 42 & 157 & 8 & 35 & 35 & 4.49 \\
 & 5 & 45 & 4 & 137 & 34 & 46 & 197 & 11 & 137 & 34 & 4.38 \\
 & 10 & 54 & 5 & 563 & 33 & 44 & 260 & 24 & 566 & 32 & 4.81 \\
 & 15 & 59 & 5 & 886 & 32 & 46 & 315 & 9 & 888 & 32 & 5.34 \\
 & 20 & 67 & 7 & 1194 & 32 & 48 & 373 & 31 & 1194 & 32 & 5.57 \\
LDAP Search & 25 & 74 & 8 & 1664 & 32 & 51 & 471 & 29 & 1667 & 32 & 6.36 \\
(binary) & 50 & 108 & 14 & 3377 & 32 & 55 & 906 & 83 & 3343 & 33 & 8.39 \\
 & 100 & 166 & 31 & 6915 & 32 & 58 & 2099 & 156 & 6913 & 32 & 12.64 \\
 & 200 & 291 & 33 & 15503 & 32 & 59 & 5785 & 245 & 15491 & 32 & 19.88 \\
 & 500 & 769 & 97 & 40316 & 32 & 60 & 27898 & 578 & 40353 & 32 & 36.28 \\
 & 606 & 973 & 169 & 48887 & 32 & 61 & 40658 & 377 & 48924 & 32 & 41.79 \\
\hline
 & 2 & 69 & 5 & 37 & 210 & 306 & 221 & 13 & 26 & 207 & 3.20 \\
 & 5 & 92 & 9 & 180 & 207 & 299 & 380 & 15 & 90 & 205 & 4.13 \\
 & 10 & 130 & 15 & 450 & 205 & 303 & 549 & 47 & 186 & 204 & 4.22 \\
 & 15 & 168 & 23 & 716 & 205 & 303 & 777 & 62 & 279 & 201 & 4.63 \\
IMS Update & 20 & 202 & 31 & 836 & 204 & 303 & 1098 & 131 & 401 & 202 & 5.44 \\
(binary) & 25 & 243 & 16 & 1047 & 204 & 303 & 1352 & 256 & 498 & 203 & 5.56 \\
 & 50 & 386 & 21 & 2201 & 203 & 304 & 3332 & 226 & 1100 & 201 & 8.63 \\
 & 100 & 701 & 100 & 3976 & 201 & 306 & 9227 & 256 & 2105 & 200 & 13.16 \\
 & 200 & 1718 & 170 & 8352 & 200 & 301 & 31009 & 218 & 4764 & 198 & 18.05 \\
\hline
\end{tabular}
}
\caption{Results of alignment experiment using six different operation types.}
\label{results.tab}
\end{table*}

%%%%%%%%%%%%%%%%%%%%%%%%%%%%%%%%%%%%%%%%%%%%%%%%%%%%%%%%%%%%%%%%%%%%%%%%%%%%

%% file: conclusions.tex
% Document Type: LaTeX
% Master File: conclusions.tex
% Author: Jean-Guy Schneider, Peter Mandile, Steve Versteeg
% Last Modified: Fri Aug 28 2015, 15:41

% Conclusions for the ASWEC 2015 submissions

%%%%%%%%%%%%%%%%%%%%%%%%%%%%%%%%%%%%%%%%%%%%%%%%%%%%%%%%%%%%%%%%%%%%%%%%%%%%

\section{Conclusions and Future Work}
\label{conc.sec}

In this paper, we have presented a novel, time and memory efficient Multiple
Sequence Alignment (MSA) algorithm based on Generalized Suffix Trees (GSTs).
We evaluated the accuracy and efficiency of the new algorithm against six
enterprise service message trace datasets, with the proposed algorithm
performing up to 50 times faster for large number of sequences than standard
MSA approaches. We also evaluated the new approach using a data set from
bioinformatics, demonstrating the adaptability of the approach to different
domains.

%%% Future work...
The core component of the {\MS} algorithm is the GST, which can be constructed
in linear time. The empirical measurements we have made support that {\MS}
scales with approximately linear complexity. Future work will conduct a more
thorough analysis of the algorithm's time and memory complexity, including for
the worst case and the average case. We will also do additional empirical
investigations of the proposed MSA algorithm on different data sets and
explore alternative strategies to identify the anchors in the splitting
process. Furthermore, the current implementation of trimming partially
overlapping multi sub-words is not optimal and improvements could lead to
further run-time gains.

%In future work, we intend to further analyze the time and memory complexity of
%the proposed MSA algorithm using different data sets as well as experiment
%with further strategies to identify the anchors in the splitting process. We
%are also interested in identifying worst-case scenarios where the
%divide-and-conquer approach of the algorithm does result in a slow run-time
%performance. Further, the current implementation of trimming partially
%overlapping multi sub-words is not optimal and improvements could lead to a
%further run-time gains.

For the majority of this work, we focused our attention on identifying the
overlapping sub-sequences of a set of messages in order to identify the
constant parts of regular expression-based message prototypes. Based on these
alignments, finding suitable patterns for the non-overlapping segments in
order to complete the regular expressions and evaluating their accuracy and
effectiveness is a topic of an ongoing investigation.

Finally, the linear time complexity for the construction of generalized suffix
trees makes them an ideal focus for further work. Particularly, we would like
to investigate whether the structural information of a GST alone is sufficient
to define a form of similarity between sequences that exhibits properties
similar to the more commonly used edit distance.

% ------------------------------------------------------------------------ %

%\section*{Acknowledgements}
%
%The authors would like to thank Steve Versteeg for his suggestions in various
%discussions about the topic matter.

%%%%%%%%%%%%%%%%%%%%%%%%%%%%%%%%%%%%%%%%%%%%%%%%%%%%%%%%%%%%%%%%%%%%%%%%%%%%